\documentclass[reprint,superscriptaddress,aps]{revtex4-1}
\usepackage{amsmath}
\usepackage{amssymb}
\usepackage{bm}
\usepackage{braket}
\usepackage{color}
\usepackage[varg]{txfonts}
\usepackage{varwidth}
\usepackage{dcolumn}
\usepackage[breaklinks,colorlinks=true,linkcolor=blue,urlcolor=cyan,citecolor=blue]{hyperref}
\usepackage{graphicx}

\begin{document}

\title{Enhancement of giant magnetoelectric effect in Ni-doped CaBaCo$_{4}$O$_{7}$}

\author{M. Gen}
\email{gen@edu.k.u-tokyo.ac.jp}
\affiliation{Department of Advanced Materials Science, University of Tokyo, Kashiwa 277-8561, Japan}
\affiliation{Institute for Solid State Physics, University of Tokyo, Kashiwa, 277-8581, Japan}

\author{A. Miyake}
\affiliation{Institute for Solid State Physics, University of Tokyo, Kashiwa, 277-8581, Japan}

\author{H. Yagiuchi}
\affiliation{Department of Advanced Materials Science, University of Tokyo, Kashiwa 277-8561, Japan}

\author{Y. Watanabe}
\affiliation{Department of Advanced Materials Science, University of Tokyo, Kashiwa 277-8561, Japan}

\author{A. Ikeda}
\affiliation{Institute for Solid State Physics, University of Tokyo, Kashiwa, 277-8581, Japan}

\author{Y. H. Matsuda}
\affiliation{Institute for Solid State Physics, University of Tokyo, Kashiwa, 277-8581, Japan}

\author{M. Tokunaga}
\affiliation{Institute for Solid State Physics, University of Tokyo, Kashiwa, 277-8581, Japan}
\affiliation{RIKEN Center for Emergent Matter Science (CEMS), Wako 351-0198, Japan}

\author{T. Arima}
\email{arima@k.u-tokyo.ac.jp}
\affiliation{Department of Advanced Materials Science, University of Tokyo, Kashiwa 277-8561, Japan}
\affiliation{RIKEN Center for Emergent Matter Science (CEMS), Wako 351-0198, Japan}

\author{Y. Tokunaga}
\affiliation{Department of Advanced Materials Science, University of Tokyo, Kashiwa 277-8561, Japan}

\begin{abstract}

The polar magnet CaBaCo$_{4}$O$_{7}$ is known to exhibit the largest magnetic-field-driven electric polarization change ($\Delta P$) associated with an antiferromagnetic (AFM)-ferrimagnetic (FIM) transition in a narrow temperature range between 62 and 69~K.
In this work, we investigate the effect of Ni doping on its multiferroic properties, by means of magnetization, electric polarization, dielectric constant, and magnetostriction measurements on single crystals of CaBaCo$_{3.9}$Ni$_{0.1}$O$_{7}$ up to 50~T.
In the doped material, two kinds of AFM phases appear below 78~K, accompanying negative $\Delta P$.
Upon the application of a magnetic field along any crystallographic axis, giant positive $\Delta P$ of up to $11 \sim 12$~mC/m$^{2}$ is observed along with an AFM-FIM transition in the whole temperature range below 78~K.
The giant magnetoelectric effect inherent in CaBaCo$_{4}$O$_{7}$ can be further enhanced just by a small amount of chemical substitution, in terms of (i) increasing the magnitude of $\Delta P$ and (ii) expanding the temperature range in which giant $\Delta P$ appears.

\end{abstract}

\date{\today}
\maketitle

\section{\label{Sec1}Introduction}

Since the discovery of the electric polarization switching associated with a magnetic phase transition in TbMnO$_{3}$ \cite{2003_Kim}, the search for spin-driven ferroelectricity and the relating magnetoelectric (ME) phenomena has been a central issue in condensed-matter physics \cite{2014_Tok}.
Spin cycloid with a net vector spin chirality, ${\mathbf S}_{i} \times {\mathbf S}_{j}$, was first considered as a major source of spin-driven ferroelectricity \cite{2005_Kat, 2006_Ser_1, 2006_Tan, 2007_Par}.
Subsequently, a variety of magnetic states, including a collinear spin structure, were found to potentially produce electric polarization, which can be explained by the $d$--$p$ hybridization \cite{2007_Ari, 2010_Mur} or the exchange striction mechanisms \cite{2006_Ser_2, 2008_Coi, 2009_Tok}.
For exploring nontrivial ME responses, frustrated spin systems are recognized as a promising playground because an external magnetic field can induce phase transitions among nearly degenerate magnetic states \cite{2011_Ari}.

\begin{figure}[b]
\centering
\includegraphics[width=\linewidth]{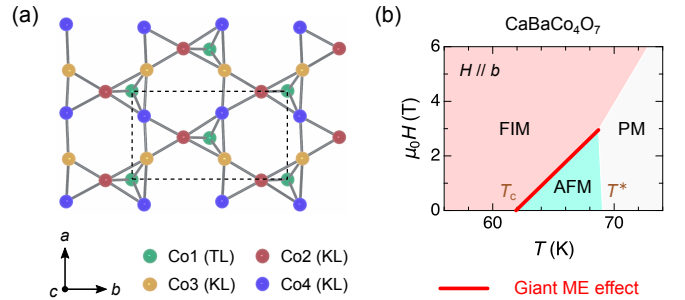}
\caption{(a) Schematic of the $c$-axis projection of the Co network in the swedenborgite lattice of CaBaCo$_{4}$O$_{7}$. The dashed rectangle represents the unit cell, which contains four inequivalent Co sites: Co1 on the triangular lattice (T) and Co2, Co3, and Co4 on the distorted kagom\'{e} lattice (K). Note that the kagom\'{e} and triangular layers are alternately stacked along the $c$-axis, and the neighboring kagom\'{e} layers alter their orientations by 180$^\circ$. For simplicity, only half of the lattice ($0 \leq z \leq 1/2$) is shown. Co atoms in the kagom\'{e} layers can be preferentially substituted by a small amount of Ni \cite{2017_Dha}. (b) Magnetic phase diagram of undoped CaBaCo$_{4}$O$_{7}$ for $H \parallel b$, illustrated based on Refs.~\cite{2012_Iwa, 2013_Cai, 2021_Cha, 2021_Omi, 2021_Omi_Doctor}. A giant $\Delta P$ is observed at the AFM-FIM phase boundary between 62 and 69~K (red line).}
\label{Fig1}
\end{figure}

The swedenborgite CaBaCo$_{4}$O$_{7}$ (CBCO), which belongs to a polar orthorhombic space group $Pbn2_{1}$, is known as a multiferroic material with magnetic frustration \cite{2009_Cai, 2010_Cai, 2011_Qu, 2012_Iwa, 2013_Cai, 2021_Cha, 2021_Omi, 2021_Omi_Doctor, 2014_Joh, 2017_Fis, 2012_Sin, 2015_Bor}.
The crystal structure is comprised of an alternate stacking of kagom\'{e} and triangular layers of CoO$_{4}$ tetrahedra,  as shown in Fig.~\ref{Fig1}(a).
The buckling distortion in the kagom\'{e} layers results in four inequivalent Co sites, and induces a charge ordering of Co2/Co3 sites with Co$^{2+}$ and Co1/Co4 sites with Co$^{3+}$(3d$^{6}$/3d$^{7}$\underline{L}) \cite{2009_Cai, 2010_Cai}.
The high-temperature magnetization measurement evidences a large negative Weiss temperature $\Theta_{\rm CW}=-890$~K and the development of an antiferromagnetic (AFM) short-range order below $\sim$360~K \cite{2011_Qu}.
On cooling, the system undergoes a transition to an AFM state at $T^{*} \approx 69$~K and finally enters a ferrimagnetic (FIM) ground state at $T_{\rm C} \approx 62$~K \cite{2012_Iwa, 2013_Cai, 2021_Cha, 2021_Omi}, where the Co2/Co3 spins form ferromagnetic zigzag chains with the easy axis along the $b$-axis and the Co1/Co4 spins point roughly antiparallel to them \cite{2010_Cai}.
The magnetic transition from the AFM to the FIM phase takes place also when applying a magnetic field along the $b$-axis in an intermediate-temperature range of $T_{\rm C} < T < T^{*}$ [Fig.~\ref{Fig1}(b)] \cite{2013_Cai, 2021_Cha}.
Remarkably, the AFM-FIM transition accompanies a giant electric polarization change of $\Delta P \approx 8$~mC/m$^{2}$ along the $c$-axis, which is the largest value reported to date \cite{2013_Cai, 2021_Cha, 2021_Omi}.
Due to the nonswitchable character of the electric polarization direction, CBCO is classified as pyroelectric \cite{2013_Cai}.
According to the previous neutron diffraction and thermal expansion measurements, the emergence of the FIM state is accompanied by an abrupt decrease in the lattice constant $c$ \cite{2010_Cai, 2013_Cai, 2021_Cha}, suggesting that the giant ME effect is attributed to the exchange striction \cite{2014_Joh, 2017_Fis}.

From the viewpoint of magnetism, the effect of partial chemical substitution on CBCO was systematically studied by many researchers \cite{2012_Sei, 2014_Sei, 2015_Aur, 2020_Loh, 2012_Sar_1, 2013_Zou, 2016_Sei, 2012_Sar_2, 2017_Yu, 2017_Dha}.
One approach is the divalent Sr$^{2+}$ substitution for the nonmagnetic Ca$^{2+}$/Ba$^{2+}$, revealing that a few percent of Sr doping significantly suppresses the FIM ordering while maintaining the orthorhombic crystal structure \cite{2014_Sei, 2015_Aur, 2020_Loh}.
This signals that the ferrimagnetism in undoped CBCO is quite sensitive to the local structural distortion.
Another approach is the substitution for the magnetic Co, and it has been pointed out that the selective substitution of sites with the target valence is possible by a small amount of doping up to $\sim$20~\% \cite{2012_Sar_1, 2013_Zou, 2016_Sei, 2012_Sar_2, 2017_Yu, 2017_Dha}.
In the case of the nonmagnetic Al$^{3+}$/Ga$^{3+}$ substitution for Co$^{3+}$, the magnetic order is statistically destroyed at the benefit of magnetic frustration even though the Co2/Co3 ferromagnetic chains remain untouched.
As a consequence, the FIM ground state is gradually suppressed and the spin- (or cluster-) glass-like behavior develops with increasing the doping level \cite{2012_Sar_1, 2013_Zou}.
Of particular interest is the divalent ion substitution for Co$^{2+}$, which directly affects the Co2/Co3 ferromagnetic chains.
In contrast to Al$^{3+}$/Ga$^{3+}$ doping, only a few percent of nonmagnetic Zn$^{2+}$ doping abruptly weakens the ferrimagnetism and instead leads to an AFM ground state, where the fragmented Co2/Co3 ferromagnetic chains would be distributed and couple antiferromagnetically with each other \cite{2012_Sar_1, 2012_Sar_2}.
Similarly, magnetic Ni$^{2+}$ doping has been also found to enhance the antiferromagnetism and brings a collinear AFM ground state \cite{2017_Dha}.

Bearing in mind that giant $\Delta P$ is linked to the AFM-FIM transition in CBCO, one can expect that Zn$^{2+}$ or Ni$^{2+}$ doping, which brings AFM ground states as mentioned above, is a promising approach to control/enhance the ME effect.
Such a study, however, has been so far limited to the polycrystalline samples of Ni-doped CBCO \cite{2017_Yu, 2017_Dha}.
In this paper, we report a single-crystal investigation of the ME effect in CaBaCo$_{3.9}$Ni$_{0.1}$O$_{7}$ (CBCNO) by means of magnetization, electric polarization, dielectric constant, and magnetostriction measurements up to 50~T.
In zero field, CBCNO undergoes successive magnetic transitions at $T_{\rm N1}=78$~K and $T_{\rm N2}=62$~K, both of which are found to accompany a negative change in $\Delta P$ along the $c$-axis.
Strikingly, giant positive $\Delta P$ of up to $11 \sim 12$~mC/m$^{2}$ is observed along with the field-induced AFM-FIM transition in the whole temperature range below $T_{\rm N1}$.
In other words, Ni doping on CBCO dramatically increases the magnitude of $\Delta P$ itself as well as expands the temperature range exhibiting giant $\Delta P$.
We also present the anisotropic nature of magnetostructural transitions and the accompanied ME responses of CBCNO in detail.

\section{\label{Sec2}Experimental methods}

Single crystals of CaBaCo$_{4-x}$Ni$_{x}$O$_{7}$ ($x=0.1$) were grown by the floating zone method under air as in Ref.~\cite{2021_Omi}.
Here, NiO was additionally mixed with the starting ingredients in stoichiometric proportions.
At the end of growth, the crystal rod was quenched down to room temperature.
Single-phase CaBaCo$_{4-x}$Ni$_{x}$O$_{7}$ was obtained near the top of the crystal rod, which was confirmed by the powder X-ray diffraction pattern on crushed single crystals.
Scanning electron microscopy-energy dispersive X-ray (SEM-EDX) analysis revealed the chemical composition of $x=0.096$, in agreement with the target value ($x=0.1$) within the measurement accuracy.

\begin{figure}[t]
\centering
\includegraphics[width=\linewidth]{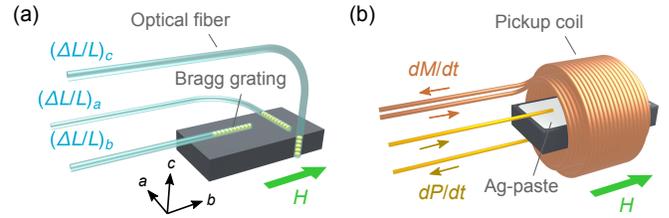}
\caption{Experimental configurations for (a) simultaneous three-axial magnetostriction/thermal expansion measurements in static magnetic fields and (b) simultaneous magnetization and electric polarization measurements in pulsed high magnetic fields. Both figures are illustrated for $H \parallel b$.}
\label{Fig2}
\end{figure}

The magnetization $M$ up to 7~T was measured using a commercial magnetometer (MPMS, Quantum Design).
The electric polarization change $\Delta P$, dielectric constant $\varepsilon'$, and magnetostriction/thermal expansion $\Delta L/L$ up to 9~T were measured using a commercial cryostat equipped with a superconducting magnet (PPMS, Quantum Design).
$\Delta P$ along the $c$-axis was obtained by integrating the pyroelectric current measured by using an electrometer (6517A, Keithley).
$\varepsilon'$ along the $c$-axis was measured at 10~kHz by using an LCR meter (E4980A, Agilent).
$\Delta L/L$ was measured by the fiber-Bragg-grating (FBG) technique using an optical sensing instrument (Hyperion si155, LUNA).
Here, the distortions along the three principal axes, $(\Delta L/L)_{a}$, $(\Delta L/L)_{b}$, and $(\Delta L/L)_{c}$, were simultaneously measured by gluing three optical fibers to one crystal sample using epoxy Stycast1266, as illustrated in Fig.~\ref{Fig2}(a).

$M$, $\Delta P$, and $\Delta L/L$ up to $\sim$50~T were measured using a non-destructive pulsed magnet ($\sim$11~ms duration for $M$ and $\Delta P$, and $\sim$36~ms duration for $\Delta L/L$) installed at the International MegaGauss Science Laboratory of the Institute for Solid State Physics, University of Tokyo, Japan    .
$M$ was measured by the conventional induction method using a coaxial pickup coil.
$\Delta P$ along the $c$-axis was obtained by integrating the pyroelectric current \cite{2007_Mit} after subtracting the 120-K data as the background.
Here, $M$ and $\Delta P$ were simultaneously measured with the setup illustrated in Fig.~\ref{Fig2}(b), making it possible to extract unambiguous information on how $\Delta P$ develops as the magnetic state changes.
The longitudinal $\Delta L/L$ was measured by the FBG technique using the optical filter method \cite{2018_Ike}.

In this work, we applied a magnetic field in three orthogonal axes.
Here, it should be noted that as-grown single crystals of CaBaCo$_{4-x}$Ni$_{x}$O$_{7}$ contain micrometer-sized trigonally-twinned domains with the common $c$-axis due to a hexagonal-to-orthorhombic structural transition at much above room temperature \cite{2017_Fis, 2021_Omi} (A polarizing microscope image of domains is shown in Fig.~S1 in Supplemental Material \cite{SM}).
Therefore, one has to take care of the contributions of three kinds of domains when applying a magnetic field along the $b$- or $a$-axis of one domain, as will be discussed in Sec.~\ref{Sec3_4}.
In the following, the field direction is simply written as ``for $H \parallel b$'' or ``for $H \parallel a$''.
We prepared three pieces of rectangular parallelpiped crystals obtained from the same crystal rod, and used Sample \#1 ($a \times b \times c = 2 \times 4 \times 1$~mm$^{3}$) for $H \parallel b$, \#2 ($4 \times 2 \times 1$~mm$^{3}$) for $H \parallel a$, and \#3 ($2 \times 1 \times 2$~mm$^{3}$) for $H \parallel c$ so as to easily perform the simultaneous measurements as mentioned above.

\section{\label{Sec3}Results and discussions}

\subsection{\label{Sec3_1}Temperature dependence of ME and dilatometric responses in zero field}

\begin{figure}[b]
\centering
\includegraphics[width=0.75\linewidth]{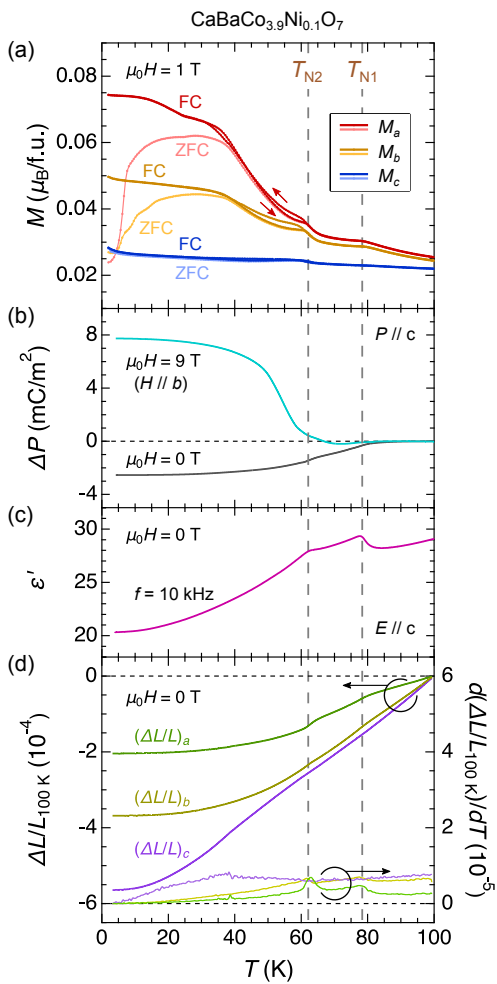}
\caption{Temperature dependence of (a) magnetization $M$ measured at 1~T for $H \parallel a$ (Sample \#2), $b$ (\#1), and $c$ (\#3) upon warming after zero-field cooling (ZFC), upon field-cooling (FC), and the following warming, (b) electric polarization change $\Delta P$ along the $c$-axis measured at zero field and 9 T for $H \parallel b$ (\#1), (c) dielectric constant $\varepsilon'$ along the $c$-axis measured at 10~kHz at zero field (\#2), and (d) thermal expansion $\Delta L/L$  and its temperature derivative along the three principal axes measured at zero field (\#2). The data in (b)--(d) were obtained upon warming after ZFC.}
\label{Fig3}
\end{figure}

Figure~\ref{Fig3} shows the temperature dependence of (a) $M$ measured at 1~T for each field direction ($M_{a}$, $M_{b}$, and $M_{c}$), (b) $\Delta P$ measured at zero field and 9 T for $H \parallel b$, (c) $\varepsilon'$ measured at zero field, and (d) $\Delta L/L$ measured at zero field for CBCNO.
At (nearly) zero field, two anomalies indicative of magnetic transitions are clearly observed at $T_{\rm N1} = 78$~K and $T_{\rm N2} = 62$~K in all the physical quantities.
These observations are almost in accordance with the previous reports on the polycrystalline samples \cite{2017_Yu, 2017_Dha}.
No hysteretic behavior can be seen at $T_{\rm N1}$ whereas a hysteresis is observed around $T_{\rm N2}$ in the $M$--$T$ curves [Fig.~\ref{Fig3}(a)], indicating that the transitions at $T_{\rm N1}$ and $T_{\rm N2}$ are of second- and first-order, respectively.
Note that $M_{a}$ and $M_{b}$ exhibit another kink around 35~K, below which the zero-field cooling (ZFC) data largely deviate from the field-cooling (FC) data [Fig.~\ref{Fig3}(a)].
These features would be attributed to the formation of short-range cluster glass introduced by magnetic frustration and randomness, as universally observed in various kinds of doped CBCO \cite{2012_Sei, 2014_Sei, 2015_Aur, 2020_Loh, 2012_Sar_1, 2013_Zou, 2016_Sei, 2012_Sar_2}.
No corresponding anomalies are observed in other physical quantities, that we will no more discuss these anomalies.

The appearance of an intermediate-temperature AFM (IM-AFM) phase between $T_{\rm N1}$ and $T_{\rm N2}$ in CBCNO is also found in undoped CBCO [Fig.~\ref{Fig1}(b)].
However, CBCNO exhibits an electric polarization change of $\Delta P = P(T_{\rm N2})-P(T_{\rm N1}) \approx -2$~mC/m$^{2}$ [Fig.~\ref{Fig3}(b)], in contrast to the undoped case with almost no change in $\Delta P$ \cite{2021_Omi}.
This difference in the ME response suggests that the magnetic structure of the IM-AFM phase in CBCNO is different from that in CBCO, which is proposed to be a noncollinear AFM state with a magnetic unit cell doubled along the $a$-axis \cite{2021_Omi}.
Another important effect of Ni doping is the ground state switching from the FIM to the AFM state, where a collinear spin configuration is proposed by powder neutron diffraction \cite{2017_Yu, 2017_Dha}.
The AFM character is reflected in the significant suppression of the magnetic moment for $H \perp c$;
on cooling from the paramagnetic state, $M$ jumps by $\sim$0.3~$\mu_{\rm B}/{\rm f.u.}$ at $T_{\rm C}$ and further increases toward low temperatures at 0.01~T for CBCO \cite{2021_Omi}, while $M$ gradually increases in the whole temperature range below $T_{\rm N1}$ and eventually reaches only less than 0.1~$\mu_{\rm B}/{\rm f.u.}$ at 1~T for CBCNO.
Here, the ratios $M_{a}/M_{c}$ and $M_{b}/M_{c}$ at low temperatures in CBCNO are much smaller than those in CBCO \cite{2012_Iwa}, indicating that the easy-plane magnetic anisotropy is weakened by Ni doping.

The ground state switching also modifies the dilatometric responses.
The previous neutron diffraction experiment on CBCO has revealed increases in the lattice constants $a$ and $b$ upon entering the FIM state \cite{2013_Cai}.
For CBCNO, in contrast, $(\Delta L/L)_{a}$ and $(\Delta L/L)_{b}$ monotonically decrease on cooling in the whole temperature range, accompanied by a tiny kink at $T_{\rm N1}$ and $T_{\rm N2}$ [Fig.~\ref{Fig3}(d)].
Although there remains an ambiguity about the actual changes in the lattice constants $a$ and $b$ due to the presence of trigonally-twinned domains, it is reasonable to conclude that both $a$ and $b$ continue to decrease upon entering the AFM state.
Moreover, $(\Delta L/L)_{c}$ does not exhibit any clear anomalies either at $T_{\rm N1}$ or at $T_{\rm N2}$ in CBCNO unlike the case of CBCO, in which an abrupt lattice shrinkage takes place along the $c$-axis upon entering the FIM state \cite{2021_Cha}.
These observations evidence that the crystal lattice in the AFM state of CBCNO is elongated along the $c$-axis compared to that in the FIM state of CBCO.

\begin{figure*}[t]
\centering
\includegraphics[width=0.85\linewidth]{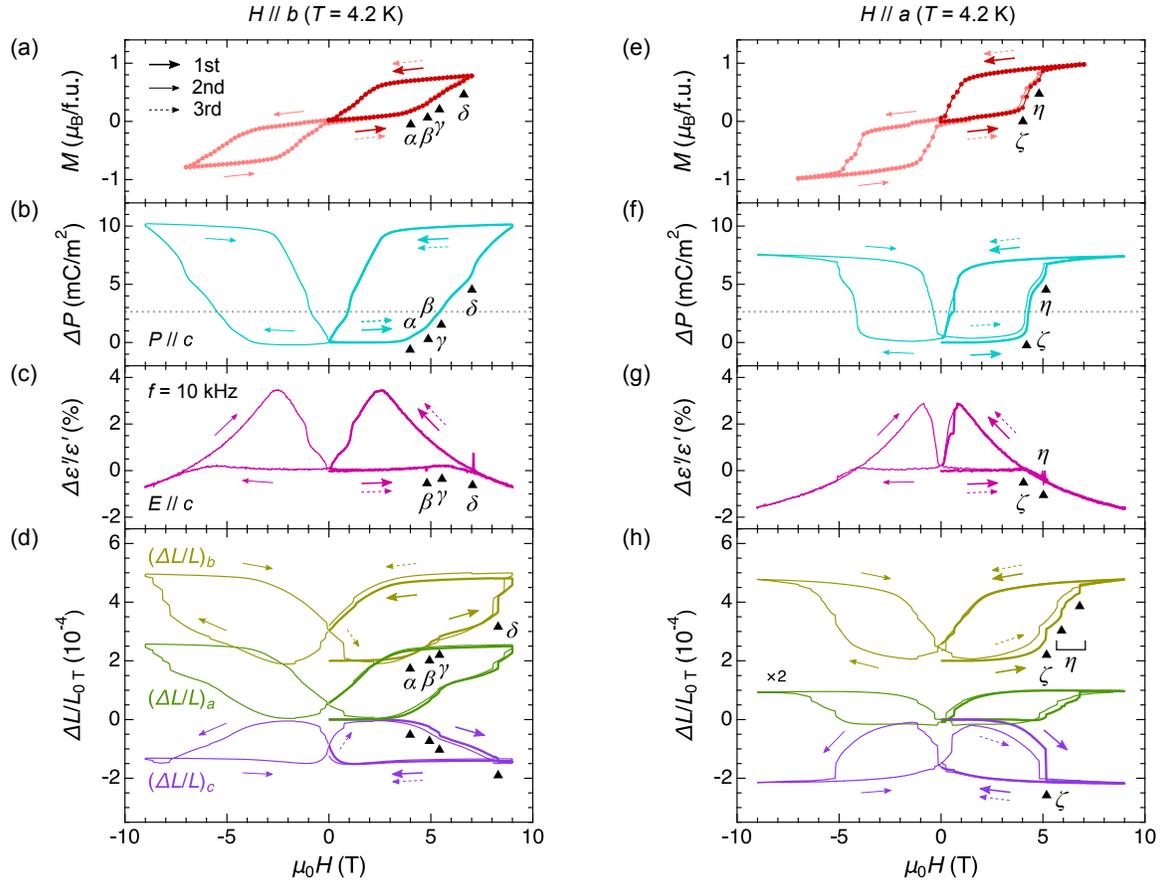}
\caption{Magnetic-field dependence of (a)(e) $M$, (b)(f) $\Delta P$ along the $c$-axis, (c)(g) $\varepsilon'$ along the $c$-axis, and (d)(h) $\Delta L/L$ along the three principal axes measured at 4.2~K in static magnetic fields for $H \parallel b$ (Sample \#1) [(a)--(d)] and $H \parallel a$ (\#2) [(e)--(h)]. The dotted lines in (b) and (f) indicate the baseline of electric polarization at 100~K in zero field. $\varepsilon'$ was measured at 10~kHz. The data of $(\Delta L/L)_{b}$ are vertically offset by $2 \times 10^{-4}$ for clarity. Arrows denote the order of field scans, as shown in the inset of (a), where the first scan is 0~T $\rightarrow$ 9~T $\rightarrow$ 0~T, the second is 0~T $\rightarrow$ $-9$~T $\rightarrow$ 0~T, and the third is 0~T $\rightarrow$ 9~T $\rightarrow$ 0~T. Each measurement was performed after zero-field cooling (ZFC). The data taken in the first field scan are displayed by thick lines. Triangles with Greek letters indicate points where an anomaly is detected in the field-increasing process of the first field scan. Here, the same letters are used for anomalies which are assigned to the identical magnetic-state change. All of these notations are omitted for $(\Delta L/L)_{a}$, which is qualitatively similar with $(\Delta L/L)_{b}$.}
\label{Fig4}
\end{figure*}

These differences highlight how the exchange striction mechanism works in the FIM state of CBCO.
In magnetic materials with a strong magnetoelastic coupling, the strain is typically related to the local spin correlator $\langle {\mathbf S}_{i} \cdot {\mathbf S}_{j} \rangle$, i.e., the relative angle between the nearest-neighbor (NN) spins \cite{2008_Zap, 2019_Ike, 2020_Miy}.
Since each kagom\'{e} layer of the Co network is connected via Co1 sites in CBCO, the strain along the $c$-axis should originate from changes in the spin correlators between Co1 and the other Co sites.
In this perspective, the crucial difference between the proposed magnetic structure in the FIM state of CBCO \cite{2010_Cai} and that in the AFM state of CBCNO \cite{2017_Dha} is in the relative angle between Co1 and Co4 spins, which are almost parallel to each other in the former while antiparallel to each other in the latter.
Thus, it can be conjectured that the driving force of the magnetostructural transition at $T_{\rm C}$ in CBCO is the stabilization of ferromagnetic spin alignments between Co1 and Co4 spins at the cost of the elastic energy, which results in the appearance of ferrimagnetism and the pronounced ME effect. 

\subsection{\label{Sec3_2} Phase transitions in static magnetic fields up to 9~T}

Figure~\ref{Fig4} shows the magnetic-field dependence of $M$, $\Delta P$, $\varepsilon'$, and $\Delta L/L$ measured at 4.2~K for $H \parallel b$ [(a)--(d)] and $H \parallel a$ [(e)--(h)].
Here, the sample was cooled down in zero field before measurements with successive field scans of (i) 0~T $\rightarrow$ 9~T $\rightarrow$ 0~T, (ii) 0~T $\rightarrow$ $-9$~T $\rightarrow$ 0~T, and (iii) 0~T $\rightarrow$ 9~T $\rightarrow$ 0~T.
In both configurations, $M$ linearly increases up to $\sim$4~T in the field-increasing process, followed by a sequence of steps indicative of metamagnetic transitions.
The $M$--$H$ loops show a butterfly shape instead of a rectangle shape as observed in CBCO \cite{2012_Iwa}, reflecting the ground state switching from the FIM to the AFM state in zero field.
Such metamagnetic transitions are not observed for $H \parallel c$ below 9~T (not shown), which is compatible with the existence of the easy-plane magnetic anisotropy.

\begin{figure*}[t]
\centering
\includegraphics[width=\linewidth]{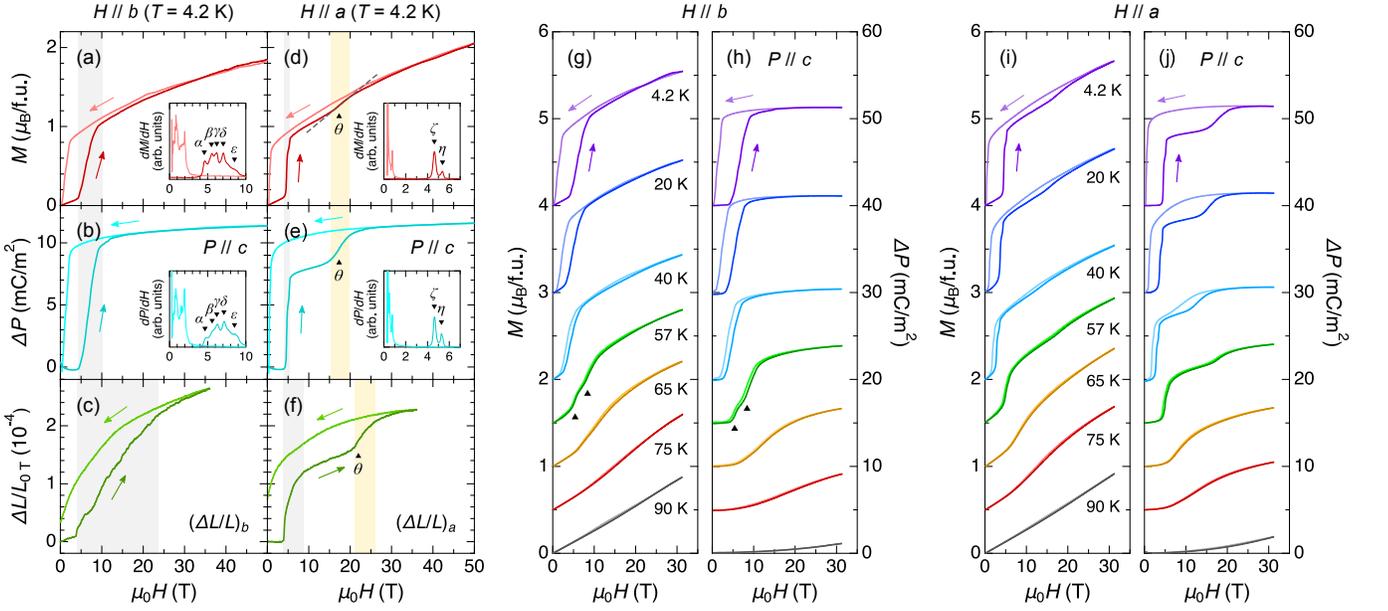}
\caption{[(a)--(f)] Magnetic-field dependence of (a)(d) $M$, (b)(e) $\Delta P$ along the $c$-axis, and (c)(f) $\Delta L/L$ along the field direction measured at 4.2~K in pulsed magnetic fields for $H \parallel b$ (Sample \#1) [(a)--(c)] and $H \parallel a$ (\#2) [(d)--(f)]. The inset of (a), (b), (d), and (e) shows the field-derivative of each physical quantity in the low-field region. Here, the definitions of Greek letters are identical with Fig.~\ref{Fig4}. Thanks to the simultaneous measurements of $M$ and $\Delta P$, the $dM/dH$ and $dP/dH$ anomalies are observed at exactly the same magnetic fields. [(g)--(j)] Magnetic-field dependence of (g)(i) $M$ and (h)(j) $\Delta P$ along the $c$-axis measured at selected temperatures for $H \parallel b$ (Sample \#1) [(g) and (h)] and $H \parallel a$ (\#2) [(i) and (j)]. The data except for 90~K are vertically offset for clarity. Triangles in (g) and (h) denote the boundaries of phase~Q [see Fig.~\ref{Fig6}(a)].}
\label{Fig5}
\end{figure*}

Importantly, an anisotropy also manifests in the character of the metamagnetic transitions for $H \parallel b$ and $H \parallel a$. 
When applying a magnetic field $H \parallel b$, $M$ gradually deviates from the linear field-dependence from $\mu_{0}H_{\alpha} \approx 4.0$~T, followed by multi-stage tiny steps at $\mu_{0}H_{\beta} \approx 4.8$~T, $\mu_{0}H_{\gamma} \approx 5.4$~T, and $\mu_{0}H_{\delta} \approx 6.6$~T [Fig.~\ref{Fig4}(a)].
These magnetic transitions are accompanied by dielectric anomalies and a clear increase in $\Delta P$, which amounts to $\sim$10~mC/m$^{2}$ at 9~T [Figs.~\ref{Fig4}(b) and \ref{Fig4}(c)].
A comparable electric polarization change of $\Delta P \approx 8$~mC/m$^{2}$ can also be induced by sweeping temperature at 9~T [Fig.~\ref{Fig3}(b)].
The observed $\Delta P$--$T$ curve at 9~T is similar with that for CBCO in zero field \cite{2021_Cha, 2021_Omi}, suggesting the emergence of a field-induced FIM state in CBCNO.
Although it is elusive how each spin is oriented with respect to the magnetic field in this field-induced FIM state, we infer that the Co2/Co3 spins are ferromagnetically aligned and the Co1/Co4 spins are antiferromagnetically coupled to them as in the zero-field FIM state of CBCO.
In the following, we assume such a state in CBCNO as the FIM state.
Notably, the magnitude of the magnetic field-driven $\Delta P$ on the AFM-FIM transition in CBCNO is enhanced compared to that in CBCO because there is an additional contribution of the negative $\Delta P$ in the initial AFM state as mentioned in Sec.~\ref{Sec3_1}.
In contrast to the case of $H \parallel b$, the metamagnetic transitions for $H \parallel a$ are rather sharp.
The $M$--$H$ and $\Delta P$--$H$ curves exhibit a jump at $\mu_{0}H_{\zeta} \approx 4.0$~T and a subsequent small jump at $\mu_{0}H_{\eta} \approx 4.8$~T in the field-increasing process for the first field scan [Figs.~\ref{Fig4}(e) and \ref{Fig4}(f)].
Note that these critical fields are a bit higher than those for the second and third field scans possibly due to the pinning effect of magnetic domains.
$\varepsilon'$ also exhibits a moderate slope change at $H_{\zeta}$ and a spike-like anomaly at $H_{\eta}$ [Fig.~\ref{Fig4}(g)], as in the case of $H \parallel b$ [Fig.~\ref{Fig4}(c)].
The metamagnetic transitions terminate just above $H_{\eta}$ and $\Delta P$ saturates at $\sim$7.5~mC/m$^{2}$, which is smaller than that observed for $H \parallel b$.

The multi-step AFM-FIM transitions in CBCNO are further supported by the magnetostrictive behaviors, exhibiting multi-step elongation along the $a$- and $b$-axes and contraction along the $c$-axis for both field directions with increasing a magnetic field above 4~T [Figs.~\ref{Fig4}(d) and \ref{Fig4}(h)].
Note that the number and/or the positions of anomalies observed in the $\Delta L/L$--$H$ curve are not completely reproducible.
This would be owing to the mechanical stress caused by the glue covering the sample surface.
From the observed $\Delta L/L$ behaviors, one can get information on how the spin configurations along each orthogonal direction statistically change across the phase transitions; e.g., for $H \parallel a$, $(\Delta L/L)_{c}$ exhibits a jump at $H_{\zeta}$ but no detectable anomalies at $H_{\eta}$ unlike $(\Delta L/L)_{a}$ and $(\Delta L/L)_{b}$ [Fig.~\ref{Fig4}(h)], signaling that at $H_{\eta}$ the magnetic structure changes only within the $ab$ plane.
Interestingly, a remanence is seen for all the measured $\Delta L/L$, and the contraction along the $a$- and $b$-axes and the elongation along the $c$-axis continue down to $\sim$$-1$~T upon reversal of a magnetic field, which is a hallmark of linear magnetostriction \cite{1994_Bor, 2017_Jai}.
This would be attributed to the presence of remnant FIM components at 0~T.
Indeed, the imperfection of the hysteresis-loop closing is also observed for the $M$--$H$ curves at 4.2~K [Figs.~\ref{Fig4}(a) and \ref{Fig4}(e)], so that piezomagnetism (and piezoelectric effect) might also be expected in this compound.

\subsection{\label{Sec3_3} Phase transitions in pulsed magnetic fields up to 50~T}

In order to obtain an overall picture of the magnetostructural transitions and the accompanied ME effects in CBCNO, we have extended the experimental investigation towards a higher magnetic-field-and-temperature ($H$--$T$) regime.
Figures~\ref{Fig5}(a)--\ref{Fig5}(f) show the magnetic-field dependence of $M$, $\Delta P$, and the longitudinal $\Delta L/L$ measured at 4.2~K in pulsed magnetic fields of up to 50~T for $H \parallel b$ and for $H \parallel a$.
All of the $M$ and $\Delta P$ jumps observed in static fields are perfectly reproduced in the pulsed-field measurements, as seen in the $dM/dH$ and $dP/dH$ peaks [insets of Figs.~\ref{Fig5}(a)(d) and \ref{Fig5}(b)(e), respectively].
For $H \parallel b$, the series of metamagnetic transitions terminate at around 10~T, where $\Delta P$ reaches $\sim$11~mC/m$^{2}$.
Then, $M$ monotonically increases between 10 and 50~T with almost no change in $\Delta P$.
$(\Delta L/L)_{b}$ also rapidly increases above 4~T, where the transitions are much broadened presumably due to the stress caused by the glue as mentioned in Sec.~\ref{Sec3_2}, making the hysteresis larger than that in the $M$--$H$ and $\Delta P$--$H$ curves [Fig.~\ref{Fig5}(c)].
For $H \parallel a$, on the other hand, another $M$ step is observed at $\mu_{0}H_{\theta} \approx 17$~T, as indicated by the dashed line in Fig.~\ref{Fig5}(d).
This $M$ step is accompanied by a further increase in $\Delta P$ by $\sim$3.5~mC/m$^{2}$, ultimately leading to almost the same magnitude of $\Delta P$ with that observed for $H \parallel b$.
As shown in Fig.~\ref{Fig5}(f), $(\Delta L/L)_{a}$ also rapidly increases across this transition (the critical field is shifted to higher), indicating that the exchange striction is responsible for the further increase in $\Delta P$ as well.
We conclude that the identical FIM state is realized above 20~T for $H \parallel b$ and $H \parallel a$, while the stabilization of an intermediate-field FIM (IM-FIM) phase in a wide field range between 5 and 15~T is the characteristic feature for $H \parallel a$.

Figures~\ref{Fig5}(g)--\ref{Fig5}(j) show the temperature evolution of the $M$--$H$ and $\Delta P$--$H$ curves for $H \parallel b$ and $H \parallel a$.
As the temperature is increased up to 40~K, each critical field shifts to a slightly lower field in the field-increasing process and the hysteresis gets smaller for both field directions.
Here, an increase in $\Delta P$ larger than 10~mC/m$^{2}$ is seen.
The hysteresis is almost closed above 57~K, while metamagnetic transitions accompanied by a large electric polarization change of $\Delta P = 5 \sim 10$~mC/m$^{2}$ still take place up to 75~K.
Recalling that CBCO exhibits a giant field-induced $\Delta P$ only in a narrow temperature range between 62 and 69~K \cite{2013_Cai, 2021_Cha, 2021_Omi} [Fig.~\ref{Fig1}(b)], the temperature range showing giant $\Delta P$ is widened by more than 10 times in CBCNO.

Furthermore, a metamagnetic transition accompanied by large $\Delta P$ of $\sim$12~mC/m$^{2}$ is also observed for $H \parallel c$ in CBCNO when applying a magnetic field higher than 20~T (Fig.~S2 in Supplemental Material \cite{SM}).
The hysteretic behaviors in the $M$--$H$ and $\Delta P$--$H$ curves and their temperature evolutions are similar to those for $H \parallel b$ and $H \parallel a$, suggesting an AFM-FIM transition also for $H \parallel c$.
To the best of our knowledge, this is the first observation of the ME effect for $H \parallel c$ in the CBCO family.

\subsection{\label{Sec3_4}Magnetic phase diagram of CaBaCo$_{3.9}$Ni$_{0.1}$O$_{7}$}

\begin{figure}[t]
\centering
\includegraphics[width=0.75\linewidth]{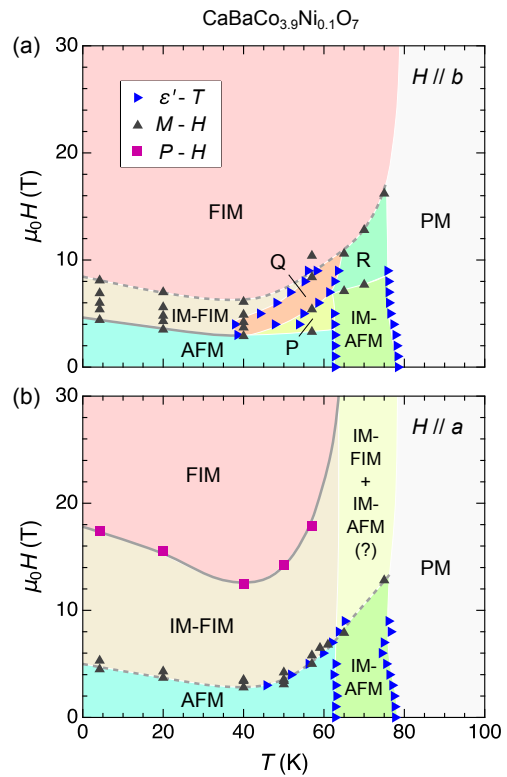}
\caption{$H$--$T$ phase diagrams of as-grown CaBaCo$_{3.9}$Ni$_{0.1}$O$_{7}$ single crystals with trigonally-twinned domains for (a) $H \parallel b$ (Sample \#1) and (b) $H \parallel a$ (\#2). Phase boundaries are determined from the dielectric anomalies observed in the warming process at constant magnetic fields (Fig.~S3 in Supplemental Material \cite{SM}) and the $dM/dH$ (or $dP/dH$) peaks observed in the field-increasing process in the pulsed-field measurements (Fig.~\ref{Fig5}). The possible AFM-FIM phase boundaries for domain I (domains II/III), defined as Fig.~\ref{Fig7}, are drawn by gray solid (dashed) lines.}
\label{Fig6}
\end{figure}

From the high-field $M$ and $\Delta P$ measurements, we construct the $H$--$T$ phase diagrams of CBCNO with trigonally-twined domains, as shown in Fig.~\ref{Fig6}.
Here, we complementarily plot phase boundaries from the temperature scans of $\varepsilon'$ measured at various magnetic fields (Fig.~S3 in Supplemental Material \cite{SM}).
For $H \parallel b$, three additional phases, P, Q, and R, which exist in a closed $H$--$T$ regime, are identified around $T_{\rm N2}$ [Fig.~\ref{Fig6}(a)]; e.g., the emergence of phase Q is clearly observed as a staircase shape in the $M$--$H$ and $\Delta P$--$H$ curves at 57~K, as indicated by triangles in Figs.~\ref{Fig5}(g) and \ref{Fig5}(h).
Such a complexity in the phase diagram compared to that of CBCO [Fig.~\ref{Fig1}(b)] should stem from the suppression of ferrimagnetism as well as the increase in the magnetic frustration caused by Ni doping, which would also be responsible for the broad multi-stage metamagnetic transitions observed at lower temperatures.
For $H \parallel a$, on the contrary, the absence of a novel phase is confirmed but instead the IM-FIM phase is found to be robust even near $T_{\rm N2}$.
Notably, the shape of the phase boundary between the IM-FIM and the FIM phases at $H_{\theta}$ is similar with that between the AFM and the IM-FIM phases at $H_{\zeta}$, satisfying the relation $H_{\theta} \approx 4 H_{\zeta}$ at all the measured temperatures below $T_{\rm N2}$.
This trend is reminiscent of the separation of identical phase transitions arising from the multi-domain nature of the crystal, as observed in GaV$_{4}$S$_{8}$ \cite{2015_Kez}.

\begin{figure}[t]
\centering
\includegraphics[width=0.85\linewidth]{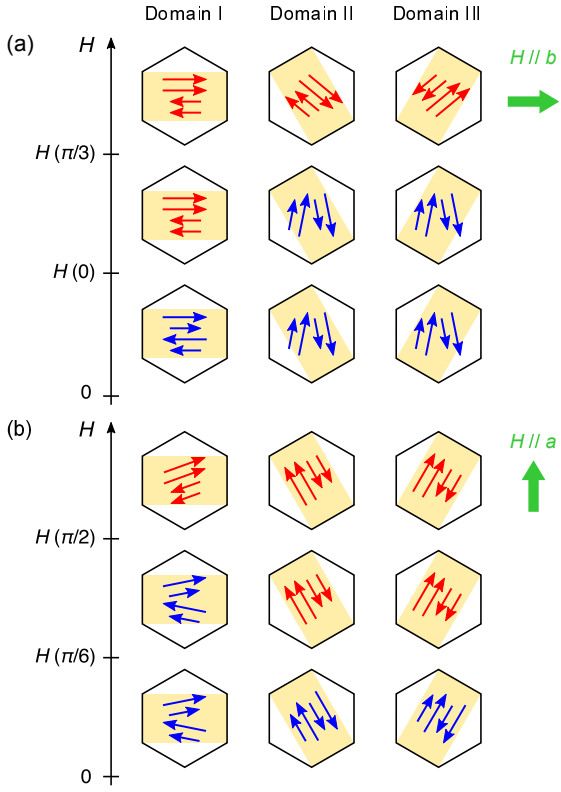}
\caption{Schematic of AFM-FIM transitions in three kinds of trigonally-twinned domains for (a) $H \parallel b$ and (b) $H \parallel a$. The critical fields are defined by Eq.~(\ref{eq1}). Long (short) arrows represent the Co1/Co4 (Co2/Co3) spins. Spin configurations drawn by blue (red) correspond to the AFM (FIM) state. Although it is unclear how each spin is oriented relative to the magnetic field, a candidate magnetic structure is depicted for each domain assuming that the magnetization easy axis is the $a$-axis and a spin-flop transition does not occur in this field region.}
\label{Fig7}
\end{figure}

In order to understand the anisotropic nature of the $H$--$T$ phase diagram for $H \parallel b$ and $H \parallel a$, we assume that (i) the crystals contain three kinds of trigonally-twinned domains (I, II, and III) with almost the same populations (see Fig.~\ref{Fig7}), (ii) each domain independently responds to an external magnetic field, and (iii) each domain does not undergo a spin-flop transition in the field region currently under consideration, but undergoes an AFM-FIM transition at a critical field
\begin{equation}
\label{eq1}
H(\phi)=\frac{H_{a}H_{b}}{\sqrt{H_{a}^2 \cos^{2}\phi + H_{b}^2 \sin^{2}\phi}}
\end{equation}
when applying a magnetic field in the $ab$ plane and at an angle of $\phi$ ($0 \leq \phi \leq \pi/2$) with $b$.
Here, $H_{a}$ and $H_{b}$ are the critical fields for $\phi = 0$ and $\pi/2$, respectively.
Under these assumptions, each domain should contribute to approximately one third of the total electric polarization change of $\Delta P \approx 11$~mC/m$^{2}$.
Bearing in mind that the critical fields in domains II and III are the same for $H \parallel b$ and $H \parallel a$, the observed $M$ step with an increase in $\Delta P$ by $\sim$3.5~mC/m$^{2}$ at a higher field $H_{\theta}$ can be attributed to the AFM-FIM transition in domain~I [Figs.~\ref{Fig5}(d) and \ref{Fig5}(e)].
This indicates that the magnetization easy axis is the $a$-axis, so that it would be reasonable to consider that the AFM spins are oriented along the $b$-axis in zero field.
Accordingly, for $H \parallel b$, the AFM-FIM transition in domain I should correspond to the onset of the series of metamagnetic transitions at $H_{\alpha}$ [Fig.~\ref{Fig5}(a)].
If we set $\mu_{0}H_{a}=\mu_{0}H_{\theta}=17$~T and $\mu_{0}H_{b}=\mu_{0}H_{\alpha}=4$~T in Eq.~(\ref{eq1}), the critical fields in domains~II/III for $H \parallel b$ and $H \parallel a$ can be estimated to $\mu_{0}H(\pi/3)=7.4$~T and $\mu_{0}H(\pi/6)=4.6$~T, respectively.
This qualitatively reproduces the main feature of the experimental results, showing significant changes in $M$ and $\Delta P$ around $6 \sim 7$~T for $H \parallel b$ and sharp metamagnetic transitions accompanied by $\Delta P \approx 7$~mC/m$^{2}$ around $4 \sim 5$~T for $H \parallel a$ (Fig.~\ref{Fig4}).
Figure~\ref{Fig7}(a) and \ref{Fig7}(b) illustrate the ways of the AFM-FIM transitions in three kinds of domains for $H \parallel b$ and $H \parallel a$, respectively.
Based on the above discussion, we draw the AFM-FIM phase boundaries for domain I (domains II/III) by gray solid (dashed) lines in Fig.~\ref{Fig6}.

The actual phase transitions in CBCNO seem more complicated.
In real samples, the magnetic state of each domain would be affected by the local stress caused by magnetostriction in the surrounding domains.
Indeed, for $H \parallel b$, $dM/dH$ and $dP/dH$ exhibit broad humps with at least five peaks over $4 \sim 10$~T at low temperatures.
Also for $H \parallel a$, the AFM-FIM transition in domains II/III is separated into two successive transitions at $H_{\zeta}$ and $H_{\eta}$.
The importance of the stress on the magnetic state is corroborated by our magnetostriction measurements, where the artificial shift of the critical fields seems to take place due to the glue covering the sample surface, as mentioned in Secs.~\ref{Sec3_2} and \ref{Sec3_3}.
A further systematic investigation of the effects of pressure or stress on the magnetic phase diagram and the ME effect in CBCNO will be also meaningful.

\section{\label{Sec4}Conclusion}

We have investigated the magnetic transitions and the ME effect of the polar magnet CaBaCo$_{3.9}$Ni$_{0.1}$O$_{7}$ up to 50~T using single crystals.
At zero field, two AFM transitions are found with a second-order transition at $T_{\rm N1}=78$~K and a first-order transition at $T_{\rm N2}=62$~K, both of which are accompanied by a dielectric anomaly and a negative change in $\Delta P$ of $\sim$$-$2~mC/m$^{2}$ along the $c$-axis.
On the application of a magnetic field along any crystallographic axes, a series of metamagnetic transitions with a giant positive change in $\Delta P$ of up to $11 \sim 12$~mC/m$^{2}$ are observed in the whole temperature range below $T_{\rm N1}$, demonstrating that the giant ME effect inherent in CBCO is enhanced.
These transitions are accompanied by multi-stage expansion in the $ab$ plane and contraction in the $c$-axis of the crystal lattice, evidencing the emergence of a field-induced FIM state.
Thus, we conclude that the enhancement of the giant ME effect is achieved by realizing the AFM ground state instead of the FIM one in low fields through Ni doping.
The doping of nonmagnetic divalent cations such as Zn$^{2+}$ would also be a promising approach to enhance the ME effect of CBCO, which has not been investigated so far.

Ferrimagnets have recently attracted growing attention for their use in spintronics devices because of the ferromagnetic-like high controllability of the net magnetization and antiferromagnetic-like fast spin dynamics \cite{2022_Kim}.
In this regards, the seek for novel physical properties and functionalities in ferrimagnets gets more important, although there remain versatile research directions to them.
Our present finding is an intriguing answer to them and potentially facilitates another discovery of the enhancement of the ME effect simply by chemical substitution in reported multiferroic materials.

\section*{Acknowledgments}

The authors appreciate for fruitful discussion to Dr. T. Omi.
This work was supported by the JSPS KAKENHI Grants-In-Aid for Scientific Research (No. 19H01835, No. 19H05826, and No. 20J10988) and Basic Science Program No. 18-001 of Tokyo Electric Power Company (TEPCO) memorial foundation.
M.G. was supported by the JSPS through a Grant-in-Aid for JSPS Fellows.


\end{document}